\documentclass[notoc,12pt,twoside,nohyper]{JHEP} 
\usepackage{epsfig}
\usepackage{amssymb}
\newcommand\fverb{\setbox\pippobox=\hbox\bgroup\verb}
\newcommand\fverbdo{\egroup\medskip\noindent%
            \fbox{\unhbox\pippobox}\ }
\newcommand\fverbit{\egroup\item[\fbox{\unhbox\pippobox}]}
\newbox\pippobox
\title{ Renormalization and factorization scale analysis of $\overline{b}b$
production in antiproton-proton collisions}

\author{by J. Ch\'{y}la
\\
    Center of Particle Physics,
    Institute of Physics AS CR, Na Slovance 2, Prague 8, Czech Republic\\
    E-mail: \email{chyla@fzu.cz}}

\abstract{There a sizable and systematic discrepancy between
experimental data on the $\overline{b}b$ production in
$\overline{\mathrm p}$p, $\gamma$p and $\gamma\gamma$ collisions
and existing theoretical calculations within perturbative QCD.
Before interpreting this discrepancy as a signal of new physics, it
is important to understand quantitatively the ambiguities of
conventional calculations. In this paper the uncertainty coming from
renormalization and factorization scale dependence of finite order
perturbation calculations of the total cross section of
$\overline{b}b$ production in $\overline{\mathrm{p}}$p collisions
is discussed in detail. It is shown that the mentioned discrepancy
is reduced significantly if these scales are fixed via the
Principle of Minimal Sensitivity.}


\begin{document}

\maketitle 

\section{Introduction}
\label{intro}
\begin{figure}[hb]\centering\unitlength=1mm
\begin{picture}(160,55)
\put(-2,0){\epsfig{file=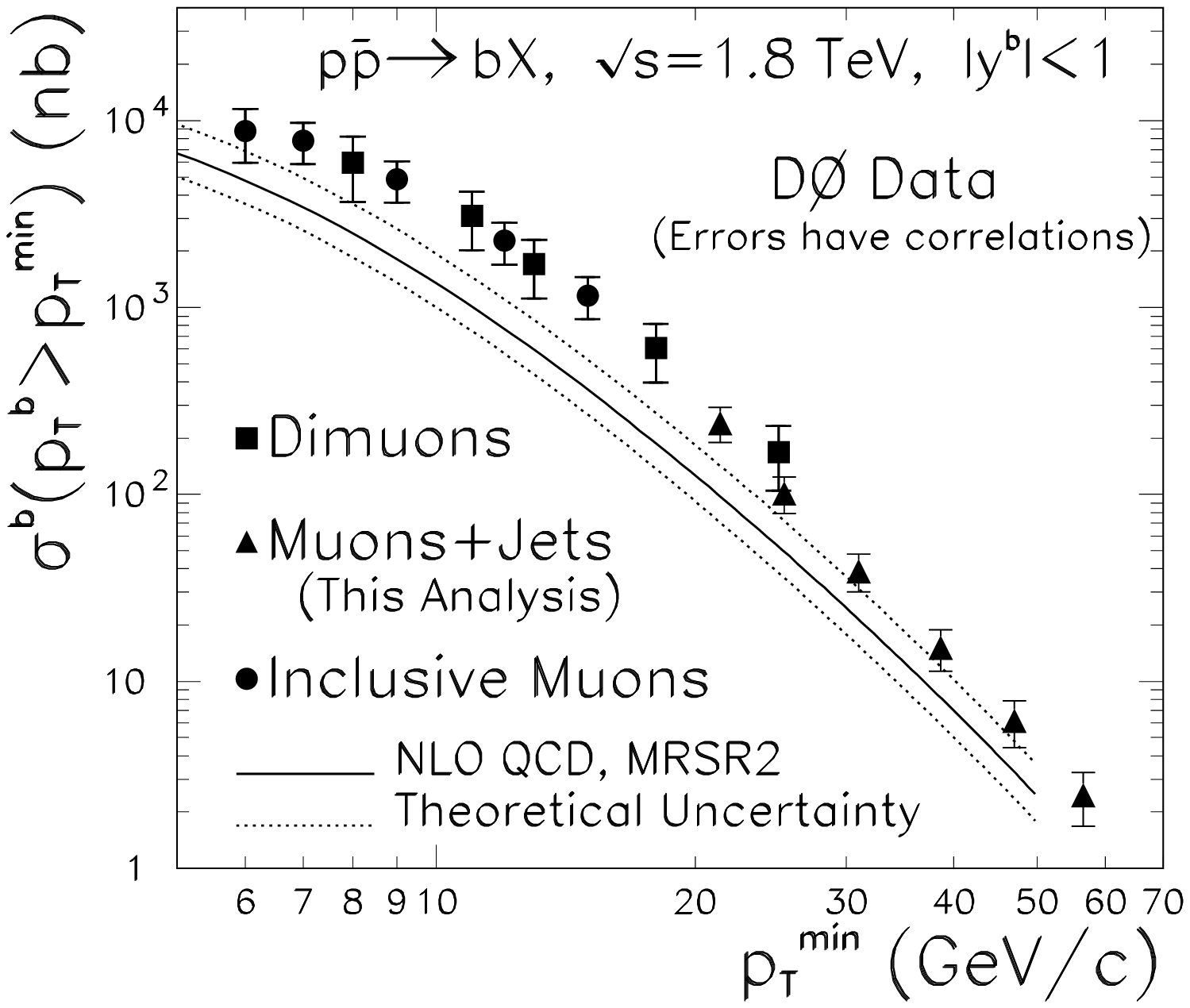, width=6.2cm}}
\put(57,-1){\epsfig{file=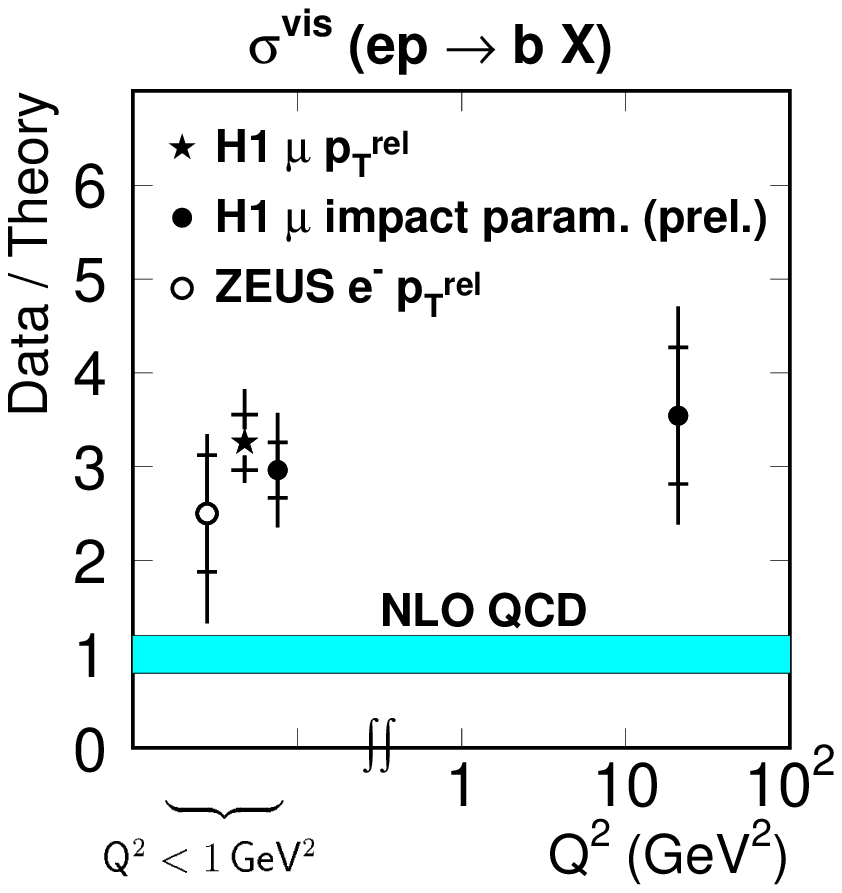,width=5.5cm}}
\put(111,3){\epsfig{file=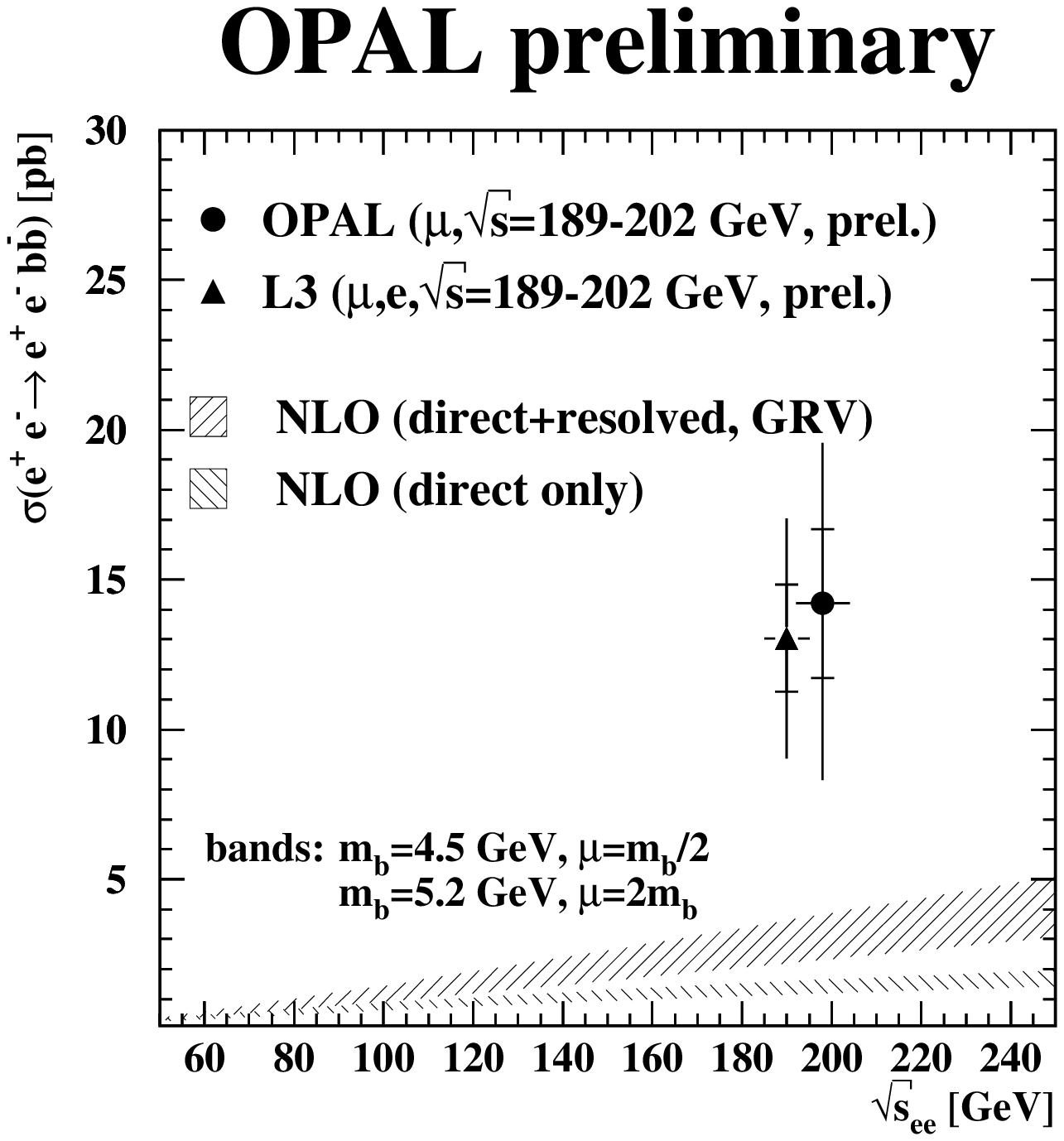,width=4.8cm}}
\end{picture}
\caption{Data on $\overline{b}b$ production in $\overline{\mathrm{p}}$p
\cite{d0b}, $\gamma$p \cite{h1b,zeusb} and $\gamma\gamma$ \cite{l3b,opalb}
collisions.}
\label{bdata}
\end{figure}

Heavy quark production in hard collisions of hadrons, leptons and photons
has been considered as a clean test of perturbative QCD. It has therefore
come as a surprise that the data on $\overline{b}b$ production in
$\overline{\mathrm{p}}$p collisions at the Tevatron \cite{d0b,cdfb},
$\gamma$p collisions at HERA \cite{h1b,zeusb} and $\gamma\gamma$
collisions at LEP2 \cite{l3b,opalb}, some of them reproduced in Fig.
\ref{bdata}, lie
systematically by a factor of about 2-4 above the median of current
theoretical calculations. For the first process the data are sufficiently
copious to measure also the differential distribution in transverse
momentum of the produced $b$ quark.
Although the data have sizable errors, this excess is too big to
be accommodated within the current calculations. The comprehensive review
of current status and related problems of the comparison of data on heavy
quark production in all three collisions with QCD calculations can be
found in \cite{frixione,nason1,nason2}.

\begin{figure}[h]\centering\unitlength=1mm
\begin{picture}(160,58)
\put(15,0){\epsfig{file=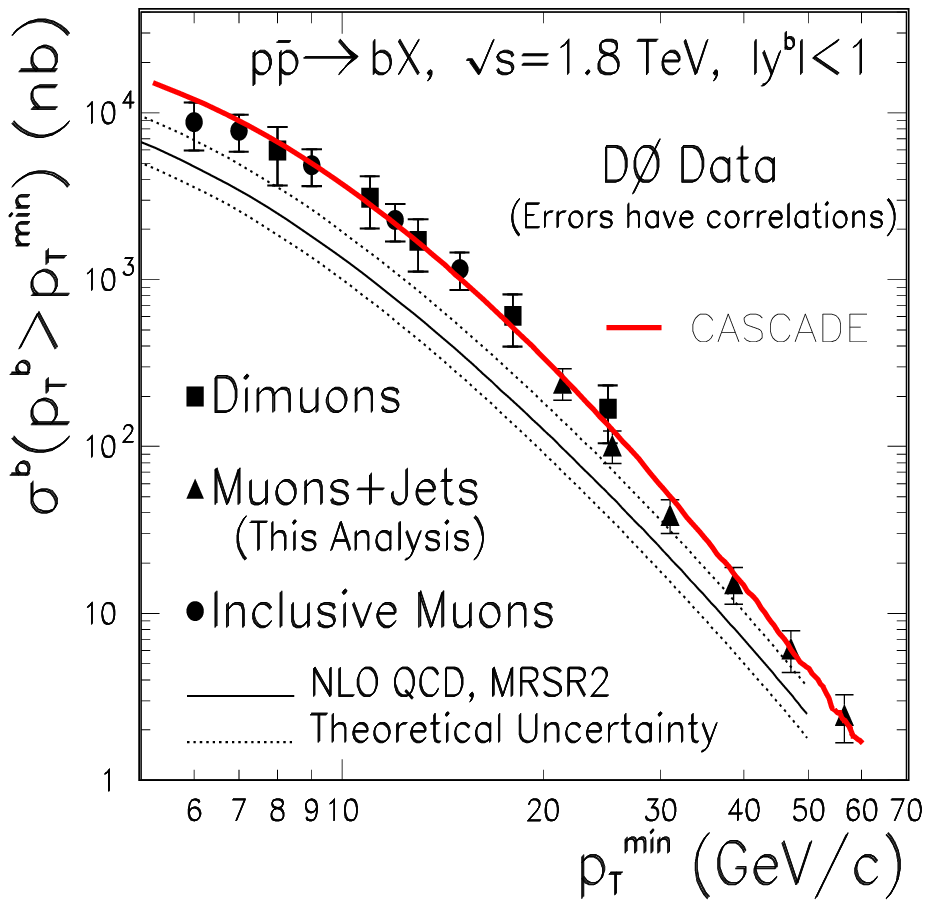, height=6cm}}
\put(80,-0.5){\epsfig{file=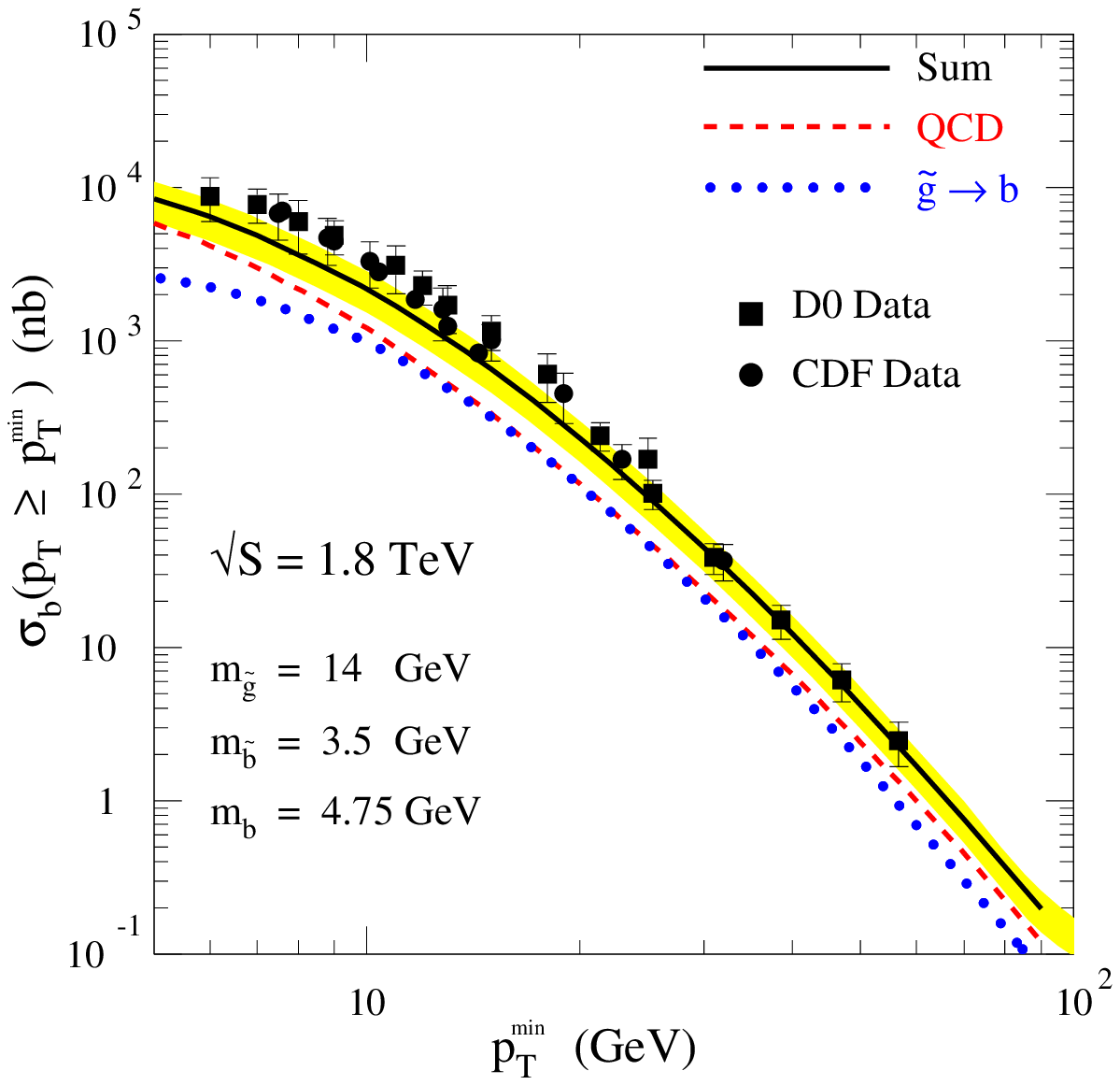,height=6cm}}
\end{picture}
\caption{D0 and CDF data on $\overline{b}b$ production compared to
calculations using the $k_T$-factorization approach as implemented in
CASCADE event generator \cite{hannes} (left) and the supersymmetry
\cite{susy} (right).}
\label{explanations}
\end{figure}

The discrepancy between the D0 and CDF data on $\overline{b}b$ production
and theoretical calculations \cite{mangano}, shown in Fig. \ref{bdata}a,
has led to suggestions that it might represent the manifestation
of effects of unintegrated gluon distribution function within the
$k_T$-factorization approach \cite{hannes,lipatov}, or even a signal of
supersymmetry \cite{susy1,susy2}. The agreement between the Tevatron data
and the calculations \cite{hannes,susy2}, reproduced in Fig.
\ref{explanations}, is, indeed, quite impressive. On the other hand,
before invoking new physics it is important to check that
the data cannot be explained by more sophisticated application of
QCD phenomenology.

The QCD calculations of heavy quark production in hard collisions of
hadrons, leptons and photons depend on a number of inputs: $\alpha_s$,
parton distribution functions (PDF) of colliding hadrons or photons,
fragmentation functions of $b$ quarks, masses of heavy quarks, and
the choice of renormalization (RS) and factorization (FS) scales
$\mu$ and $M$. Quite recently, proper parameterization of fragmentation
function has been argued \cite{nason2} to be one of the possible
sources of the disagreement of the CDF data \cite{cdfb} with NLO
QCD calculations. There are also attempts to go beyond fixed order
perturbative calculations by resumming the effects of large logs of
the type $\ln(S/4m_b^2)$ \cite{catani} or $\ln(p_T/m_b)$
\cite{cacciari}.

In this paper I investigate in detail the dependence of existing
fixed order (LO and NLO) QCD calculations \cite{htheory,guido} of the
total cross section $\sigma_{tot}(\overline{b}b)$ in
$\overline{\mathrm{p}}$p collisions on the choices of the renormalization
and factorization scales. Similar analysis of $\overline{b}b$ production
in $\gamma$p and $\gamma\gamma$ collisions will be presented elsewhere
\footnote{For discussion of the specific features of $\overline{b}b$
production in $\gamma\gamma$ collisions see \cite{ascona}}.

The paper is organized as follows. In the next Section basic facts
relevant for further discussion are collected. In Section 3
some of the methods for selecting the renormalization and factorization
scales are recalled and their relevance for the process under study
discussed. This is followed in Section 4 by the discussion of the general
form of the renormalization and factorization scale dependence of
$\sigma_{tot}(\overline{b}b)$ at the NLO. Numerical results relevant
for energies up to the Tevatron range are presented in Section 5 and
conclusions drawn in Section 6.

\section{Basic facts and formulae}
\label{basics}
The basic quantity of perturbative QCD
calculations, the renormalized color coupling $\alpha_s(\mu)$, depends on
the renormalization scale $\mu$ in a way governed by the equation
\begin{equation}
\frac{{\mathrm d}\alpha_s(\mu)}{{\mathrm d}\ln \mu^2}\equiv
\beta(\alpha_s(\mu))=
-\frac{\beta_0}{4\pi}\alpha_s^2(\mu)-
\frac{\beta_1}{16\pi^2}
\alpha_s^3(\mu)+\cdots,
\label{RG}
\end{equation}
where for $n_f$ massless quarks $\beta_0=11-2n_f/3$ and
$\beta_1=102-38n_f/3$. The solutions of (\ref{RG}) depend beside $\mu$
also on the renormalization scheme (RS). At the NLO (i.e. taking
into account first two terms in (\ref{RG})) this RS can be
specified, for instance, via the parameter $\Lambda_{\mathrm{RS}}$,
corresponding to the renormalization scale for which $\alpha_s$ diverges
\footnote{At higher orders the $\beta$-function coefficients
$\beta_k,k\ge 2$ may be used in addition to $\Lambda_{\mathrm{RS}}$ to
fix the RS.}. The
coupling $\alpha_(\mu)$ is then given as the solution of the equation
\begin{equation}
\frac{\beta_0}{4\pi}\ln\left(\frac{\mu^2}{\Lambda^2_{\mathrm{RS}}}\right)=
\frac{1}{\alpha_s(\mu)}+
c\ln\frac{c\alpha_s(\mu)}{1+c\alpha_s(\mu)},
\label{equation}
\end{equation}
where $c=\beta_1/(4\pi\beta_0)$. At the NLO the coupling $\alpha_s$ is
thus a function of the ratio $\mu/\Lambda_{\mathrm{RS}}$
and the variation of the RS for fixed scale $\mu$ is therefore equivalent
to the variation of $\mu$ for fixed RS. To vary both the renormalization
scale and scheme is legitimate, but redundant. Let me emphasize that the
choice of the RS is as important as the choice of renormalization scale.
The fact that the coupling $\alpha_s(\mu)$, as well the coefficients of
perturbation expansions, depend for fixed $\mu$ on the RS also implies
that the existence of a ``natural'' physical scale in the problem, like
the masses of heavy quarks in our case, is actually of no help in
specifying unambiguously the NLO calculation. I will come back to this
point in the next Section. If not stated otherwise, I will work in the
conventional $\overline{\mathrm{MS}}$ RS and vary the renormalization
scale $\mu$ only.

For hadrons the factorization scale dependence of PDF is determined by
the system of evolution equations for quark singlet, nonsinglet and
gluon distribution functions
\begin{eqnarray}
\frac{{\mathrm d}\Sigma(M)}{{\mathrm d}\ln M^2}& =&
P_{qq}(M)\otimes \Sigma(M)+ P_{qG}(M)\otimes G(M),
\label{Sigmaevolution}
\\ \frac{{\mathrm d}G(M)}{{\mathrm d}\ln M^2} & =&
P_{Gq}(M)\otimes \Sigma(M)+ P_{GG}(M)\otimes G(M), \label{Gevolution} \\
\frac{{\mathrm d}q_{\mathrm {NS}}(M)}{{\mathrm d}\ln M^2}& =&
P_{\mathrm {NS}}(M)\otimes q_{\mathrm{NS}}(M),
\label{NSevolution}
\end{eqnarray}
where
\begin{eqnarray}
\Sigma(x,M) & \equiv &
\sum_{i=1}^{n_f} \left(q_i(x,M)+\overline{q}_i(x,M)\right),
\label{singlet}\\
q_{\mathrm{NS},i}(x,M)& \equiv &(q_i(x,M)+\overline{q}_i(x,M))-
\frac{1}{n_f}\Sigma(x,M),~\forall i.
\label{nonsinglet}
\end{eqnarray}
The splitting functions admit expansion in powers of $\alpha_s(M)$
\begin{equation}
P_{ij}(x,M)=\frac{\alpha_s(M)}{2\pi}P^{(0)}_{ij}(x) +
\left(\frac{\alpha_s(M)}{2\pi}\right)^2 P_{ij}^{(1)}(x)+\cdots,
\label{splitpij}
\end{equation}
where $P^{(0)}_{ij}(x)$ are {\em unique}, whereas all higher order
splitting functions $P^{(j)}_{kl},j\ge 1$ depend on the choice of the
factorization scheme (FS). Conversely, they can be taken as defining
the FS, similarly as the higher order $\beta$-function coefficients
$\beta_i,i\ge 2$ in (\ref{RG}) define, together with
$\Lambda_{\mathrm{RS}}$, the renormalization scheme. The equations
(\ref{Sigmaevolution}-\ref{NSevolution}) can be recast into evolution
equations for $q_i(x,M),\overline{q}_i(x,M)$ and $G(x,M)$.

In all calculations of this exploratory study I took $n_f=4$ and solved
the RG equation (\ref{equation}) numerically in order to guarantee
correct behaviour of its solution for small renormalization scales.

\section{Renormalization and factorization scales and their choice}
\label{scales}
Let us first recall the situation for processes involving
the renormalization scale only. For them the QCD contribution to
corresponding cross sections has the form ($a_s\equiv \alpha_s/\pi$)
\begin{equation}
r(Q)=a_s^k(\mu,{\mathrm{RS}})\left(r_0(Q)+
r_1(Q/\mu,{\mathrm{RS}})a_s(\mu,{\mathrm{RS}})+
r_2(Q/\mu,{\mathrm{RS}})a_s^2(\mu,{\mathrm{RS}})\cdots\right).
\label{pert}
\end{equation}
As an example, take the familiar ratio
\begin{equation}
R_{\tau}\equiv \frac{\Gamma(\tau^-\rightarrow \nu_{\tau}+
{\mathrm{hadrons}})}
{\Gamma(\tau^-\rightarrow \nu_{\tau}\mu^-\overline{\nu}_{\mu})}=
3\left(1+r_{\tau}\right),
\label{Rtau}
\end{equation}
where the QCD correction $r_{\tau}$ is given by the expression
(\ref{pert}) with $k=1$ and $r_0=1$. Theoretical consistency of the
expansion (\ref{pert}) implies
\begin{equation}
r^{(N)}\equiv \sum_{j=k}^{N}r_{j-k}a_s^{j}~\Rightarrow~
\frac{{\mathrm{d}}r^{(N)}}{{\mathrm{d}}\ln \mu}=O(a_s^{N+1}),
\label{consistency}
\end{equation}
i.e. the derivative of the $N$-th order approximant $r^{(N)}$ with
respect to $\ln\mu$ is of higher order than the appoximant
$r^{(N)}$ itself. This requirement determines the dependence of
the coefficients $r_k$ on $\mu$. For $r_1$ we get, for instance,
\begin{equation}
r_1(Q/\mu,{\mathrm{RS}})=kb\ln\frac{\mu}{Q}+r_1(1,{\mathrm{RS}}    )=
kb\ln\frac{\mu}{\Lambda_{\mathrm{RS}}}-\rho(Q),
\label{r1}
\end{equation}
where $\rho(Q)\equiv kb\ln(Q/\Lambda_{\mathrm{RS}})-
r_1(1,{\mathrm{RS}})$ is a renormalization scale and scheme
invariant.

There are two general approaches to selecting the renormalization scale
applicable in such circumstances: the Principle of Minimal Sensitivity
\cite{pms} (PMS) and
the method based on the concept of Effective Charges \cite{ech} (EC).
In the first approach the emphasize is put on the stability of the
results with respect to variations of $\mu$. In the absence of information
on higher order perturbative terms in (\ref{pert}) the PMS approach is
natural as it selects the point $\mu$ where the truncated
perturbation expansion is most stable and has thus locally the property
possessed by the all order result globally. The EC approach is based on
the criterion of apparent convergence of perturbation expansions and
the renormalization scale $\mu$ (and at higher orders other free
parameters) is chosen in such a way that all higher order contributions
vanish, i.e. demanding $r^{(N)}=r^{(k)}$ for all $N$.
Whatever the choice of the renormalization scale $\mu$ scale we make, it
is, however, certainly useful to investigate the stability of the
results with respect to the variations of $\mu$.

In a fixed RS, both PMS and EC approaches can be phrased in geometrical
terms as methods of selecting on a curve defined by the expression
$r^{(N)}(\mu)$ some preferred point. The PMS approach selects the
stationary point(s), whereas the EC method selects the point(s) given
by the intersection of $r^{(N)}(\mu)$ with the curve corresponding to
the lowest nontrivial order.
\begin{figure}\centering
\epsfig{file=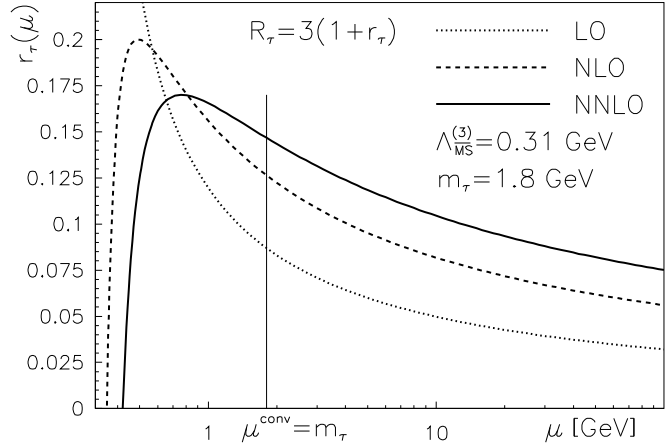,width=9cm}
\caption{Renormalization scale dependence of the ratio $r_{\tau}$ in the
$\overline{\mathrm{MS}}$ RS.}
\label{taufig}
\end{figure}
In Fig. \ref{taufig} the $\mu$ dependence of the first three approximants
$r^{(N)}_{\tau}, N=1,2,3$ for the specific case of the quantity $r_{\tau}$
defined in (\ref{Rtau}) is plotted in $\overline{\mathrm{MS}}$ RS taking
$\Lambda^{(3)}_{\overline{\mathrm{MS}}}=0.31$ GeV.
Several features of this figure are worth noting.
\begin{itemize}
\item The conventional NLO approximation in the
$\overline{\mathrm{MS}}$ RS, corresponding
to the choice $\mu_{\overline{\mathrm{MS}}}=m_{\tau}=1.8$ GeV,
gives substantially smaller result than that corresponding to the
stationary point.
\item Working in other RS, the same choice
$\mu_{\mathrm{RS}}=m_{\tau}=1.8$ GeV would give a different result
for $r_{\tau}(m_{\tau})$. In fact, \underline{any point} on the NLO
and NNLO curves can be obtained by setting
$\mu_{\overline{\mathrm{MS}}}=m_{\tau}$ in appropriately chosen RS!
The usual argument for using the $\overline{\mathrm{MS}}$ RS is
that the associated $\alpha_s$ incorporates $\ln 4\pi-\gamma_E$, the
artefact of dimensional regularization, and thus leads to smaller
expansion coefficients $r_k$ than those of MS RS. But if the rate of
convergence of perturbation expansions of physical quantities is
the main criterion for the appropriate choice of the RS, than that
based on the EC approach, rather than the $\overline{\mathrm{MS}}$
should be selected.
\item The NLO and NNLO approximants $r^{(2)}_{\tau}$ and
$r^{(3)}_{\tau}$ exhibit maxima, defining the PMS scales
$\mu^{\mathrm{PMS}}_{\mathrm{\overline{\mathrm{MS}}}}$
and the corresponding PMS optimized results $r^{(N)}_{\mathrm{PMS}}(Q)$.
No physical meaning should, however, be attached to numerical values
of $\mu^{\mathrm{PMS}}_{\overline{\mathrm{MS}}}$ as in different RS the
same maximum occurs at different scale
\footnote{For instance, $\mu^{\mathrm{PMS}}_{\mathrm{MS}}=0.38
\mu^{\mathrm{PMS}}_{\overline{\mathrm{MS}}}$, whereas
$\mu^{\mathrm{PMS}}_{\mathrm{MOM}}=2.17
\mu^{\mathrm{PMS}}_{\overline{\mathrm{MS}}}$.}.
\item At both NLO and NNLO the PMS scales
$\mu^{\mathrm{PMS}}_{\overline{\mathrm{MS}}}$ turn out to be
quite close the scales $\mu^{\mathrm{EC}}_{\overline{\mathrm{MS}}}$
chosen by the EC approach. This fact holds, with some
modification, for more complicated quantities as well.
\item
Although the preferred scales
$\mu_{\overline{\mathrm{MS}}}^{\mathrm{PMS}}$ and
$\mu_{\overline{\mathrm{MS}}}^{\mathrm{EC}}$ depend on the chosen RS,
the PMS and EC optimized results $r^{(N)}_{\mathrm{PMS}}(Q)$ and
$r^{(N)}_{\mathrm{EC}}(Q)$ do not! This fact follows directly
from the geometrical meaning of the preferred points and is crucial
for the uniqueness of the obtained predictions.
\end{itemize}
At higher orders and for more complicated physical quantities the
situation is, however, not always so simple as sketched above.
In some cases there is no genuine stationary point in the physical
region and one must then look for some other measure of ``most stable''
region, or, on the contrary, there may be more stationary points to
choose from. For physical quantities with two scales the general
criterion of the EC approach does not lead to a unique
point, but rather the curve $\mu(M)$ along which the higher order
contributions vanish. However, even in these circumstances the PMS and
EC approaches provide useful guidance
to fixing the scales and thus defining the QCD predictions.

I now come to the central aspect of the choice of the renormalization
and factorization scales. As emphasized by Politzer
\cite{politzer}, there is no compelling reason for identifying these
two scales. The point is that whereas the renormalization scale $\mu$
defines, roughly speaking, the lower limit on the virtualities of loop
particles included in the definition of the renormalized coupling, the
factorization scale $M$ specifies the upper limit on virtualities of
partons included in the definition of dressed PDF. In other words, the
renormalization scale reflects ambiguity in the treatment of short
distances, whereas the factorization scale comes from similar ambiguity
concerning large distances. It is therefore natural to keep the
renormalization and factorization scales as independent free parameters
of any finite order perturbative approximations. The calculations
presented in the next two Sections demonstrate that at the NLO
the cross section $\sigma_{tot}(\overline{b}b;M,\mu)$ depends, indeed,
on these two scales in quite different way.

Identifying these two scales may actually lead to misleading
conclusions concerning the stability of perturbative expressions.
It may act in both ways, faking the ``stability'' where there is is fact
none, but also hiding the genuine stability region when it lies away
from the diagonal point $\mu=M$. At the NLO there is an additional degree
of freedom (in fact an infinite number of them) related to the choice of
the factorization scheme, specified by the higher order splitting
functions $P_{ij}^{(1)}$. I will not exploit this freedom and
throughout this paper stay within the conventional $\overline{\mathrm{MS}}$
FS.

\section{General form of $\sigma_{tot}(Q\overline{Q})$}
\label{tevatron}
According to the factorization theorem the total cross section of
heavy quark pair production in $\overline{\mathrm{p}}$p collisions at
the center of mass energy $\sqrt{S}$ has the form
\begin{equation}
\sigma_{tot}(\overline{p}p\rightarrow\overline{Q}Q,S)=
\int\!\!\!\!\int\!\!{\mathrm{d}x}{\mathrm{d}}y
\sum_{ij}D_i^{\overline{p}}(x,M)D_j^{p}(y,M)
\sigma_{ij}(s=xyS,M),
\label{general}
\end{equation}
where partonic cross sections
$\sigma_{ij}(s,M)\equiv \sigma(ij\rightarrow\overline{b}b,s,M)$, as well
as PDF of the beam particles, $D_j^{\overline{p}}(M),D_j^p(M)$, depend
on the factorization scale $M$. The theoretical calculations of
heavy quark productions in refs. \cite{frixione,htheory,guido} are given
in terms of the heavy quark pole mass, i.e. $m_Q$ appearing in the
expressions for $\sigma_{ij}$ is just a fixed number.

The expression (\ref{general}) is basically of nonperturbative nature.
Fixed order perturbation theory enters if we insert into (\ref{general})
the solutions of the evolution equations (\ref{Sigmaevolution}-
\ref{NSevolution}) with the splitting functions $P_{ij}$ truncated to
finite order (\ref{splitpij}) and calculate partonic cross sections
$\sigma_{ij}(s,M)$ as power expansion in the coupling $\alpha_s(\mu)$
\begin{equation}
\sigma_{ij}(s,M)=\alpha_s^2(\mu)\sigma_{ij}^{(2)}(s)
+\alpha_s^3(\mu)\sigma_{ij}^{(3)}(s,M,\mu)+\cdots,
\label{expansion}
\end{equation}
taken at the renormalization scale $\mu$, which, as emphasized above, is
in general unrelated to the factorization scale $M$. Summed to all orders
of $\alpha_s$, the
r.h.s of (\ref{expansion}) is actually independent of $\mu$, whereas even
the all-order sum of (\ref{expansion}) depends on the factorization
scale $M$. The latter dependence is cancelled by that of the PDF
provided the splitting functions $P_{ij}$ in the evolution equations
(\ref{Sigmaevolution}-\ref{NSevolution}) are taken to all orders.
This difference reflects the different roles played by the renormalization
and factorization scales in expressions for physical quantities.
If only finite order approximantions in (\ref{expansion}) and
(\ref{splitpij}) are used, as is in practice always the case, the
partonic cross sections $\sigma_{ij}(s,M)$ become also $\mu$-dependent
and the expression (\ref{general}) thus becomes a function of
both $\mu$ and $M$.

At the NLO, i.e. taking into account the first two terms in
(\ref{expansion}) and (\ref{splitpij}), we get (dropping for brevity
the dependence on $S$ and specification of $\overline{b}b$ final state)
\begin{displaymath}
\sigma_{tot}^{\mathrm{NLO}}(M,\mu)=
\alpha_s^2(\mu)\left\{\int\!\!\!\!\int{\mathrm{d}}x{\mathrm{d}}y
\sum_{i=1}^{2n_f}q_i(x,M)q_i(y,M)
\left[\sigma_{q\overline{q}}^{(2)}(xy)+\alpha_s(\mu)
\sigma_{q\overline{q}}^{(3)}(xy,M,\mu)\right]+\right.
\end{displaymath}
\begin{equation}
\hspace*{0cm}2\int\!\!\!\!\int\!
{\mathrm{d}}x{\mathrm{d}}y\Sigma(x,M)G(y,M )
\alpha_s(\mu)\sigma_{qG}^{(3)}(xy,M)+
\label{pbarpnlo}
\end{equation}
\begin{displaymath}
\hspace*{3cm}\left.
\int\!\!\!\!\int\!{\mathrm{d}}x{\mathrm{d}}y G(x,M)G(y,M)
\left[\sigma_{GG}^{(2)}(xy)+\alpha_s(\mu)\sigma_{GG}^{(3)}(xy,M,\mu)
\right]\right\},
\end{displaymath}
where the sum is understood to run over $n_f$ quarks and
antiquarks, and the relation between PDF of protons and antiprotons
was taken into account.

Theoretical consistency of (\ref{general}) implies, similarly to
(\ref{consistency}), that the derivative of any finite order
approximation thereof with respect to $\ln M^2$ is of higher order
in $\alpha_s$ than such approximation itself. Performing this
derivative for (\ref{pbarpnlo}) we get
\begin{eqnarray}
\frac{{\mathrm{d}}
\sigma_{tot}^{\mathrm{NLO}}(M,\mu)}{{\mathrm{d}}\ln M^2}&=&
\int\!\!\!\!\int\!{\mathrm{d}}x{\mathrm{d}}y
G(x,M)G(y,M)W_{GG}(xy,M,\mu)\!+ \label{pbarpderivace}\\
& &\hspace*{-2cm}\int\!\!\!\!\int\!{\mathrm{d}}x{\mathrm{d}}y
\left[ \sum_{i=1}^{2n_f}q_i(x,M)q_i(y,M)W_{qq}(xy,M,\mu)
+\Sigma(x,M)G(y,M)W_{qG}(xy,M,\mu) \right].
\nonumber
\end{eqnarray}
Using (\ref{Sigmaevolution}-\ref{NSevolution}) and denoting
$\dot{f}\equiv \mathrm{d}f/\mathrm{d}\ln M^2$, the functions $W_{ij}$
are given as
\begin{eqnarray}
\!\!W_{GG}(x,M,\mu)&\!=\!& \frac{\alpha_s^3(\mu)}{\pi}
\left\{2\pi\dot{\sigma}_{GG}^{(3)}(x)+\!\!\int\!\!
{\mathrm{d}}zP^{(0)}_{GG}(z)\sigma_{GG}^{(2)}(xz)\right\}+\cdots
\label{WGGp}\\
W_{q\overline{q}}(x,M,\mu)&\!=\! & \frac{\alpha_s^3(\mu)}{\pi}
\left\{2\pi\dot{\sigma}_{q\overline{q}}^{(3)}(x)+2\!\!\int\!\!
{\mathrm{d}}zP^{(0)}_{qq}(z)\sigma_{q\overline{q}}^{(2)}(xz)\right\}
+\cdots\label{Wqqp}\\
W_{qG}(x,M,\mu)&\! =\! & \frac{\alpha_s^3(\mu)}{\pi}
\left\{2\pi\dot{\sigma}_{qG}^{(3)}(x)+\!\!\int\!\!
{\mathrm{d}}z\!\left[P^{(0)}_{qG}(z)\sigma_{q\overline{q}}^{(2)}(xz)+
P^{(0)}_{Gq}(z)\sigma_{GG}^{(2)}(xz)\right]\!\right\}\!+\cdots
\label{WqGp}
\end{eqnarray}
Theoretical consistency of (\ref{pbarpnlo}) requires that the
expressions standing in (\ref{WGGp}-\ref{WqGp}) by $\alpha_s^3$
vanish which, indeed, they do. For practical calculations it is
suitable to rewrite the expression (\ref{general}) as an integral
\begin{equation}
\sigma_{tot}(\overline{Q}Q,S)=
\sum_{ij}\int{\mathrm{d}z}\Phi_{ij}(z,M)\sigma_{ij}(s=zS,M)
\label{integral}
\end{equation}
over the product of parton fluxes
\begin{equation}
\Phi_{ij}(z,M)\equiv\int\!\!\!\int{\mathrm{d}x}{\mathrm{d}}y
D_i^{\overline{p}}(x,M)D_j^p(y,M)\delta(z-xy)
\label{fluxes}
\end{equation}
and partonic cross sections $\sigma_{ij}(z,M)$, both of which depend on the
product of fractions $z\equiv xy$ only. Because the expressions for
$\sigma_{ij}^{(3)}(s,M)$ as given in \cite{htheory} correspond to $\mu=M$,
I have restored their separate dependence on $\mu$ and $M$ by adding to
$\sigma_{ij}^{(3)}(s,M)$ the term $2\beta_0\sigma_{ij}^{(2)}\ln(\mu^2/M^2)$.

\section{Results}
\begin{figure}
\epsfig{file=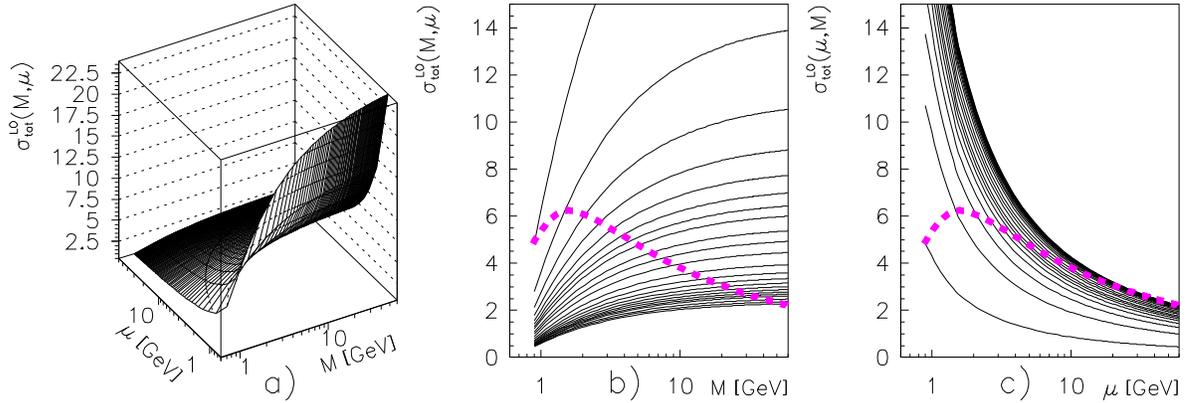,width=\textwidth}
\caption{Graphical representation of
$\sigma_{tot}^{\mathrm{LO}}(M,\mu)$ for $\sqrt{S}=630$ GeV: a) a two
dimensional surface, together with its both projections, plotted for several
values of $\mu$ in b) and $M$ in c).}
\label{fake}
\end{figure}
In Fig. \ref{fake} the LO approximation $\sigma_{tot}^{\mathrm{LO}}(M,\mu)$
i.e. the sum of the terms proportional to $\alpha_s^2$
in (\ref{pbarpnlo}), is plotted for $\sqrt{S}=630$ GeV as a two-dimensional
surface $\sigma_{tot}^{\mathrm{LO}}(M,\mu)$, as well as a function of $M$
for fixed $\mu$, and as a function of $\mu$ for fixed $M$.
There is no stability region of $\sigma_{tot}^{\mathrm{LO}}(M,\mu)$
in the sense of a stationary point on a two-dimensional surface. Keeping
one of the scales fixed and varying the other (solid curves in Figs.
\ref{fake}b-c) leads in both cases to monotonous dependence on the latter.
If, however, we set $\mu=M$, which means cutting the surface in Fig.
\ref{fake}a along the diagonal line, we obtain the dashed curve shown in
both projections in the Figs. \ref{fake}b-c, which has a spurious
stationary point at about $\mu=M\doteq 1.5$ GeV, illustrating the point
made at the end of Section 3.
\begin{figure}[h]\centering \unitlength=1mm
\begin{picture}(160,116)
\put(10,60){\epsfig{file=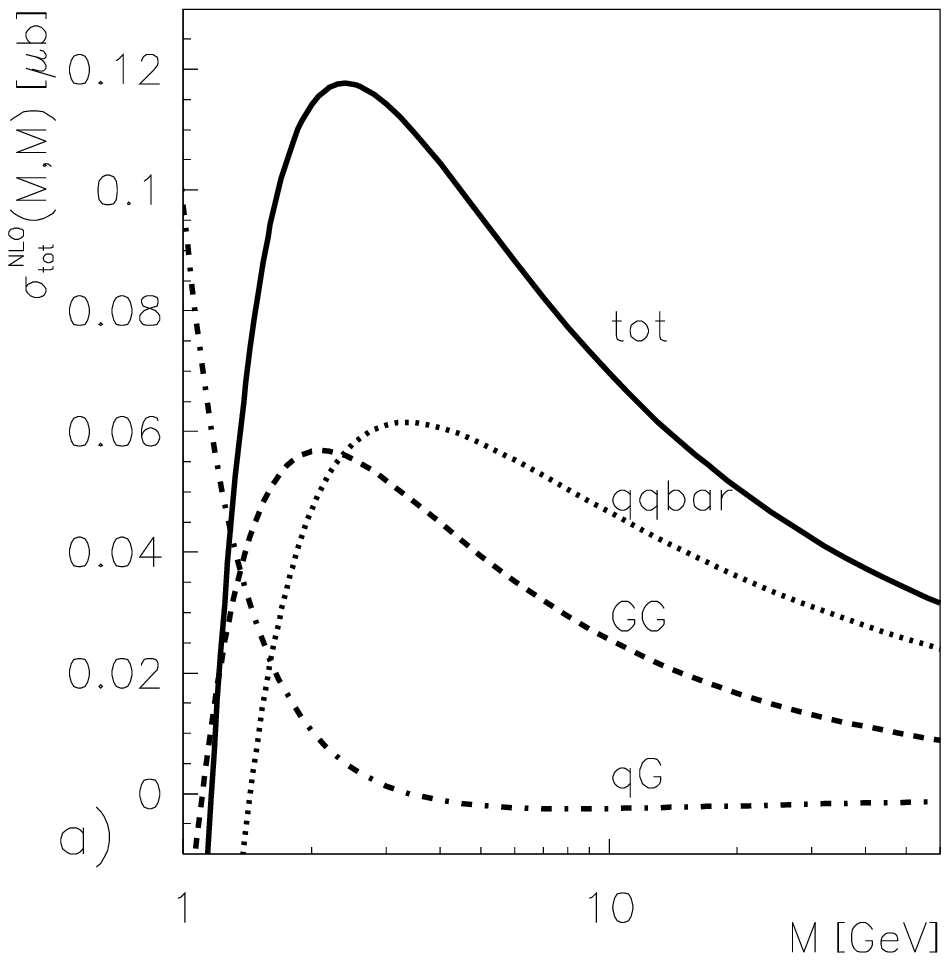,height=6cm}}
\put(70,60){\epsfig{file=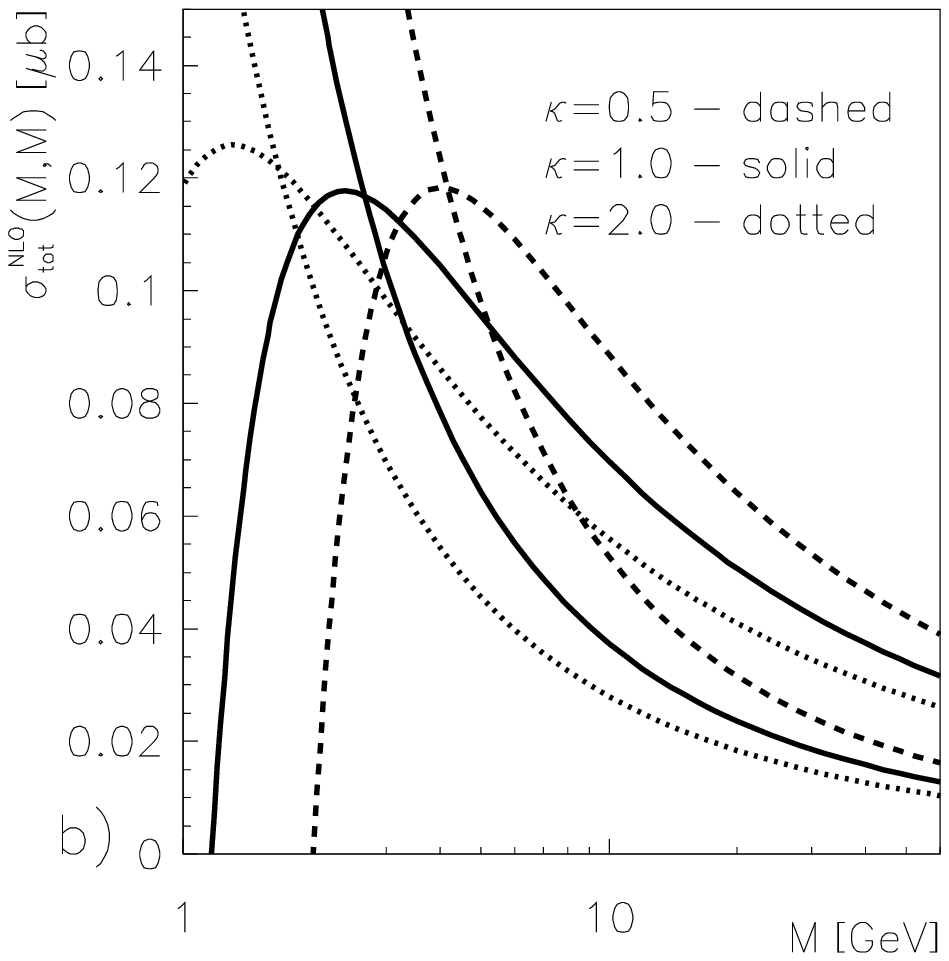,height=6cm}}
\put(10,0){\epsfig{file=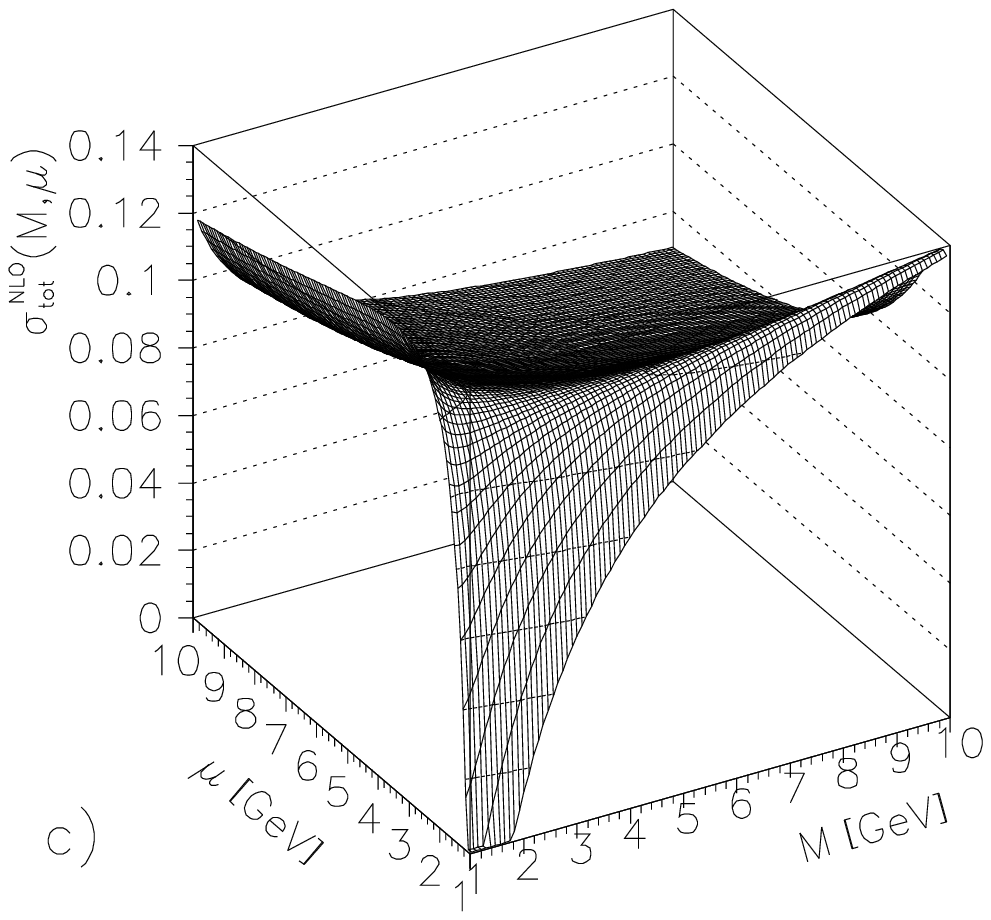,height=6cm}}
\put(70,0){\epsfig{file=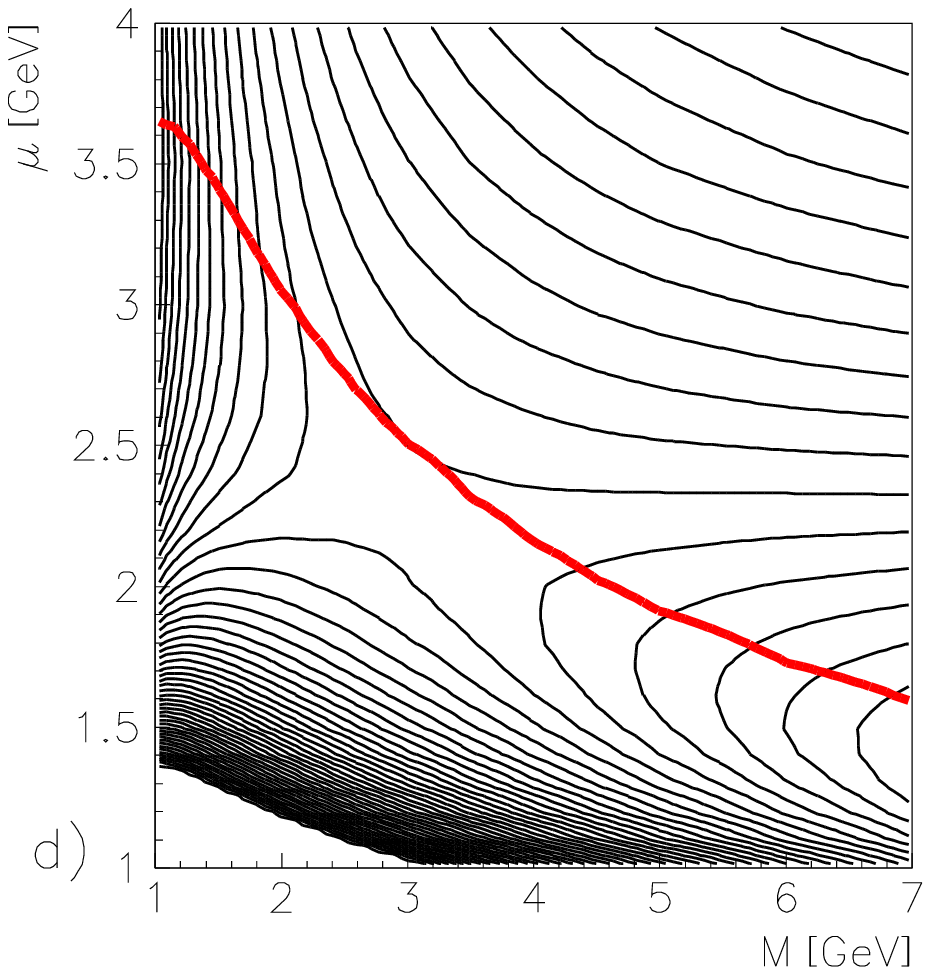,height=6cm}}
\end{picture}
\caption{Graphical representation of $\sigma_{tot}^{\mathrm{NLO}}(M,\mu)$
for $\sqrt{S}=62$ GeV.}
\label{f62}
\end{figure}

The NLO approximant $\sigma_{tot}^{\mathrm{NLO}}(M,\mu)$
is represented graphically in Fig. \ref{f62} for $\sqrt{S}=62$ GeV, one of
the energies investigated in \cite{guido}, $m_b=5$
GeV and using GRV HO PDF. Setting $\mu=M$ we get the curves in Fig.
\ref{f62}a, where the individual contributions of gluon-gluon,
quark-antiquark and quark-gluon channels are plotted as well.
The quark-antiquark channel dominates for $M\gtrsim 2.5$ GeV, but the
gluon-gluon one contributes significantly, too. At small $\mu=M$ the
quark-quark contribution turns negative, as does eventually the
gluon-gluon one, whereas that of the quark-gluon channel rises. The
latter channel gives negative contribution at large values of $M$. To
quantify the effects of the conventional assumption $\mu=M$,
$\sigma_{tot}^{\mathrm{NLO}}(M,\mu=\kappa M)$ is plotted in Fig.
\ref{f62}b for three values of $\kappa=0.5,1,2$. The position
of the stationary point as well as the corresponding value of
$\sigma_{tot}^{\mathrm{NLO}}(M,\mu=\kappa M)$, depend on the value of
$\kappa$, though not dramatically. Note that the positions of the maxima
in Fig. \ref{f62}b occur close to the intersections of the LO and NLO
curves, defining the EC predictions. The spread between the three NLO
curves at $M=m_b$ may be used as one of the measures of theoretical
uncertainty of the conventional NLO calculations. The standard method
of estimating this uncertainty is to set $M=\mu$ and vary this common
scale within the interval $(m_b/2,2m_b)$. Note, however, that the
resulting band of values of
$\sigma_{tot}^{\mathrm{NLO}}(\kappa m_b,\kappa m_b)$ depends on the
choice of RS!

\begin{figure}[h]\centering \unitlength=1mm
\begin{picture}(160,52)
\put(-2,0){\epsfig{file=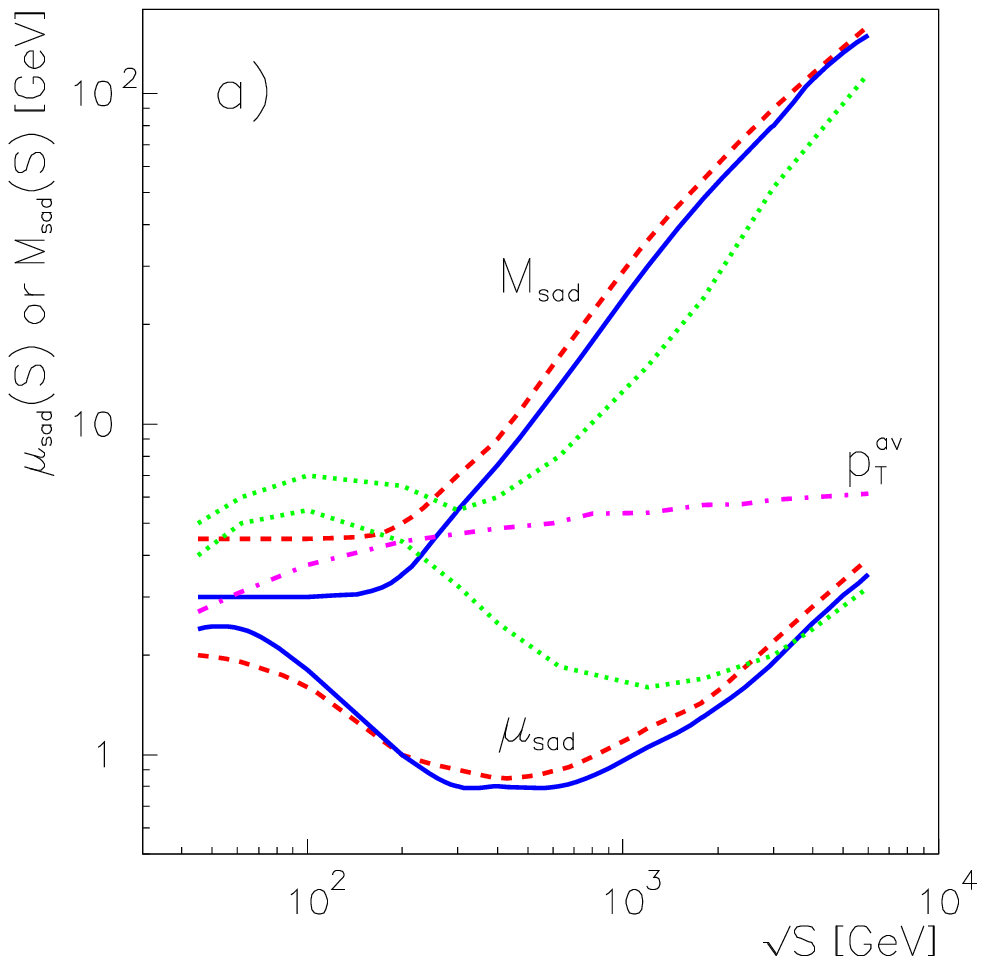,height=5.5cm}}
\put(52,0){\epsfig{file=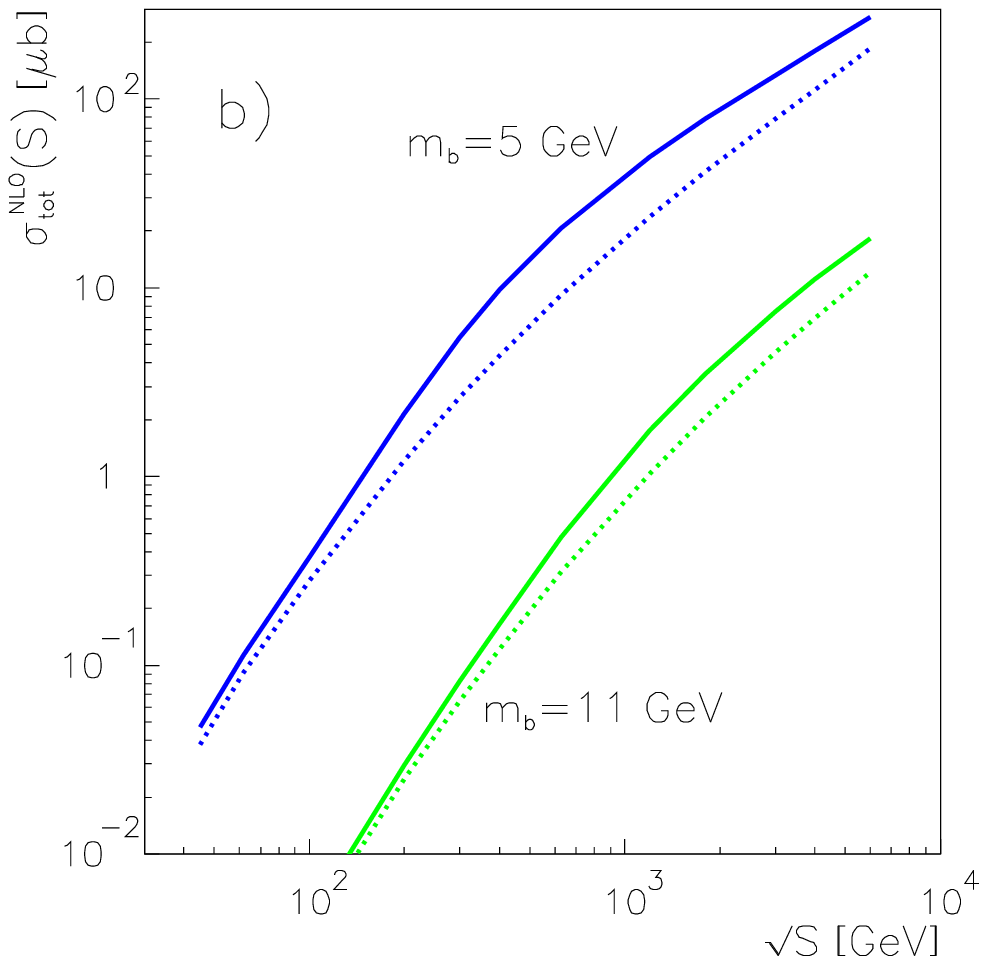,height=5.5cm}}
\put(106,0){\epsfig{file=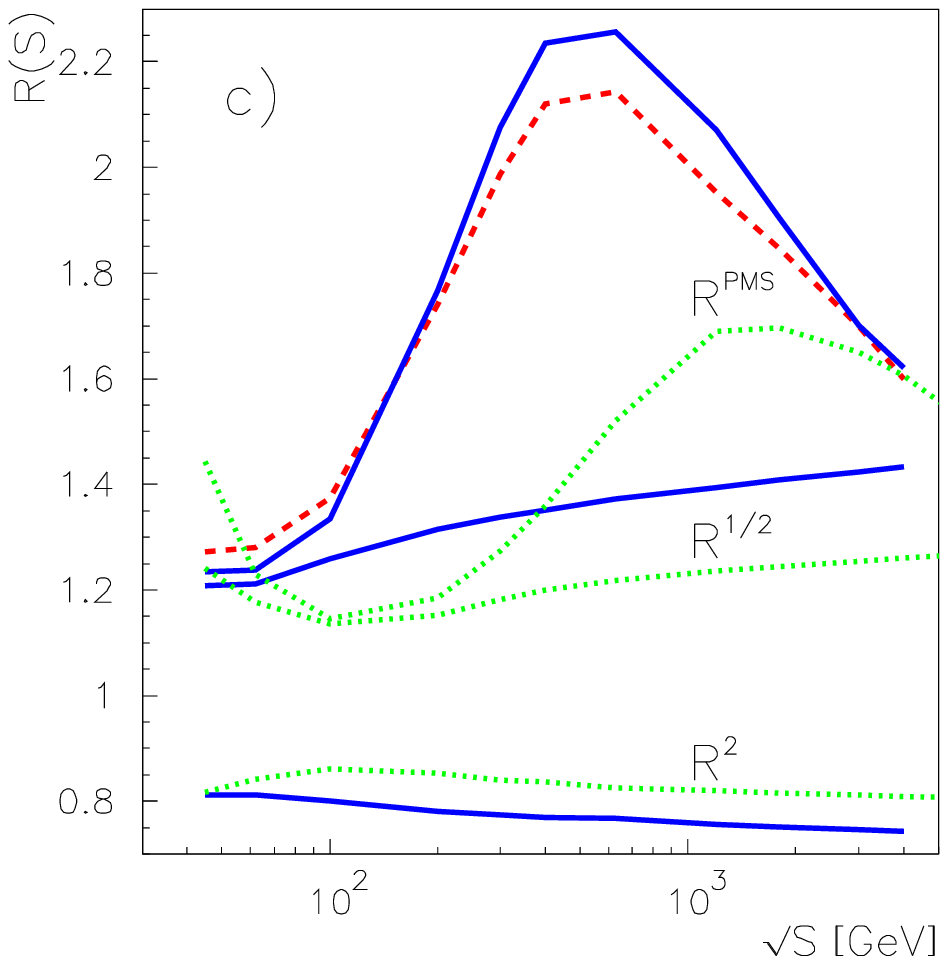,height=5.5cm}}
\end{picture}
\caption{a) Energy dependence of the position of the saddle point
of $\sigma_{tot}^{\mathrm{NLO}}(b\overline{b})$ for $m_b=5$ GeV
using GRV (solid curves) and MRS (dashed curves)
PDF of the proton. The dotted curves show the same but for
$m_b=11$ GeV. The dash-dotted curve shows the energy dependence of
the average $p_T$ of the $b$-quark.
b) The energy dependence of
$\sigma_{tot}^{\mathrm{NLO}}(b\overline{b})$ in the conventional
$\overline{\mathrm{MS}}$ RS (dotted curves) as well as the
PMS-optimized prediction (solid curves), for two values of
$m_b$.
c) The ratia $R^{1/2}$, $R^{2}$ and $R^{\mathrm{PMS}}$ for
$m_b=5$ GeV using the GRV (solid curves) and MRS (dashed curve)
PDF. The dotted curves correspond to GRV PDF and $m_b=11$ GeV.}
\label{plane}
\end{figure}

Keeping $\mu$ and $M$ as independent parameters and representing
$\sigma_{tot}^{\mathrm{NLO}}(M,\mu)$ as two-dimensional surface in Fig.
\ref{f62}c or by the corresponding contour plot in Fig. \ref{f62}d,
reveals the saddle point located approximately at $\mu=2.4$ GeV,
$M=3$ GeV. This saddle point, which defines the PMS result, lies close
to the curve in Fig. \ref{f62}d along which
$\sigma^{\mathrm{NLO}}_{tot}(M,\mu)=\sigma^{\mathrm{LO}}_{tot}(M,\mu)$,
defining the results preferred by the EC approach. Contrary to the PMS
the Effective Charges approach does not lead to a
unique prediction, but the whole range of values.
At $\sqrt{S}=62$ GeV the saddle point lies also close to the point
$\mu=M=m_b/2=2.5$ GeV, which is conventionally taken as the lower limit
on the variation of the common scale $\mu=M=m_b$.
\begin{figure}[h]\centering \unitlength=1mm
\begin{picture}(160,60)
\put(20,0){\epsfig{file=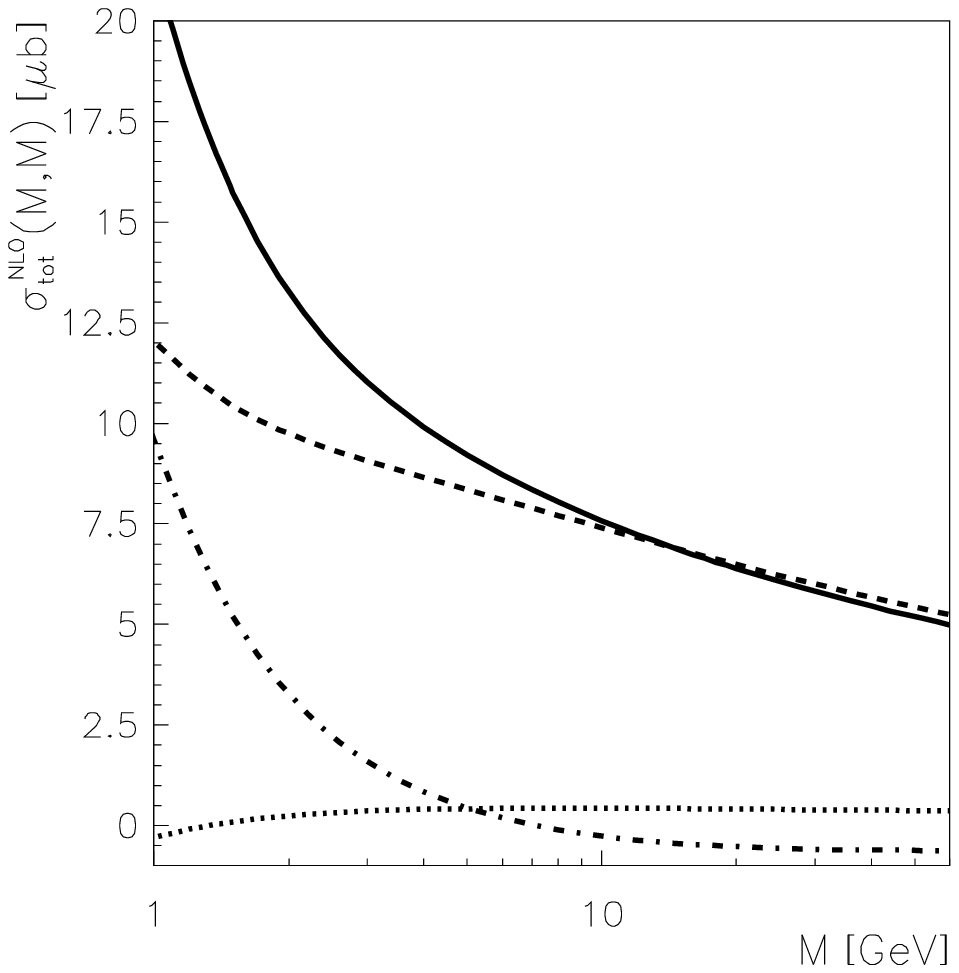,height=6cm}}
\put(80,0){\epsfig{file=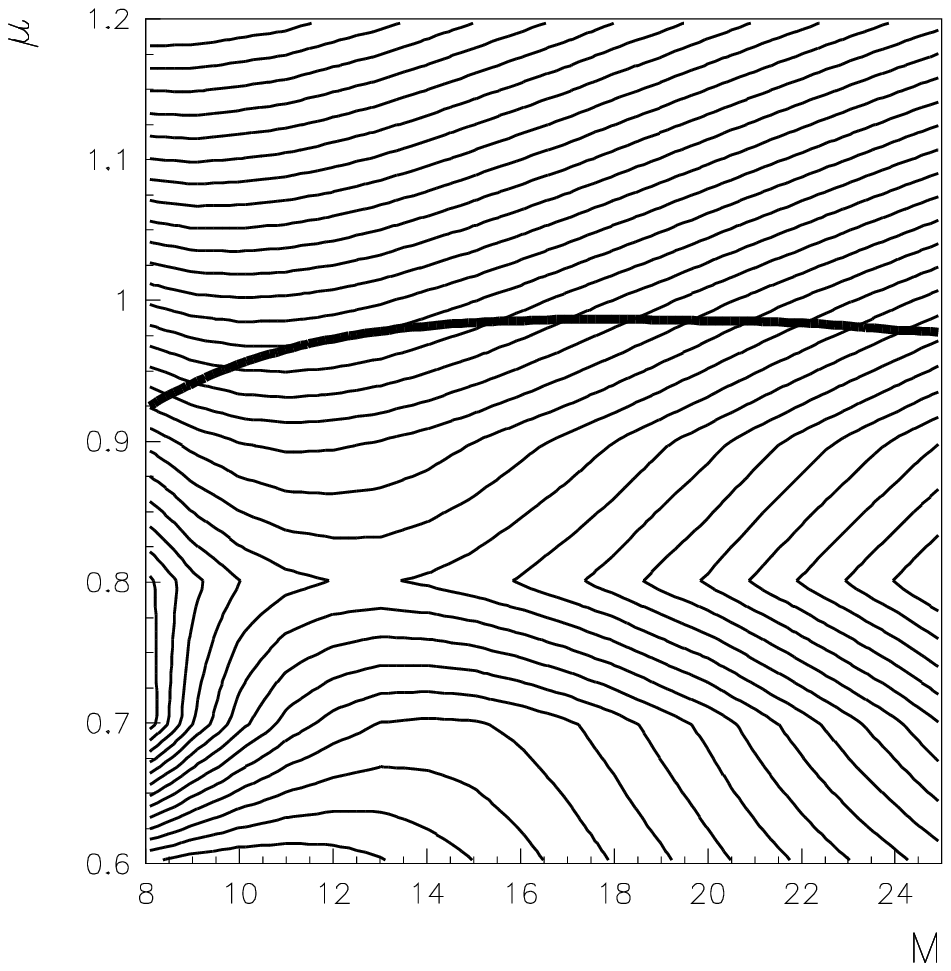,height=6cm}}
\end{picture}
\caption{The same as in Fig. \ref{f62}a,d but for
$\sqrt{S}=630$ GeV.}
\label{f630}
\end{figure}
As the collision energy increases, the situation changes
in two respects:
\begin{itemize}
\item The relative contribution of the quark-antiquark channel
decreases, its dominance at low energies being taken over by that of
the gluon-gluon channel. The gluon-quark channel grows in importance
at low values of $M$, roughly $M\lesssim 2$ GeV.
\item The position of the saddle point moves in a highly
``nondiagonal'' manner away from the line $\mu=M$ and thus away
from the conventional choice $\mu=M$. However, at all energies the
saddle point stays close to the curve defining the EC results.
\end{itemize}
These changes are illustrated in Figs. \ref{f630}, which shows the
same plots as those in Fig. \ref{f62}a,d but for $\sqrt{S}=630$ GeV.
The energy dependence of the position of the saddle point
$(\mu_{sad}(S),M_{sad}(S))$, represented by solid curves in Fig.
\ref{plane}a, illustrates yet again the
different role of renormalization and factorization scales. Whereas
$M_{sad}(S)$ stays constant up to about $\sqrt{S}\simeq 200$ GeV and then
rises roughly as some power of $S$, $\mu_{sad}(S)$ first dips down to
$\mu_{sad}\simeq 0.8$ GeV at about $\sqrt{S}\simeq 500$ GeV,
before starting to rise at about the same rate as $M_{sad}(S)$ but
remaining far below it. This large disparity between the values
of renormalization and factorization scales at the saddle point
provides the main reason for keeping these two scale as independent
free parameters.

The ``nondiagonal'' position of the saddle point at higher energies
leads to increasing difference between the PMS-optimized and
conventional $\overline{\mathrm{MS}}$ results for
$\sigma_{tot}^{\mathrm{NLO}}(S)$. This is illustrated in Fig.
\ref{plane}b which shows that the former rises more steeply in the
region $100 \lesssim \sqrt{s}\lesssim 500$ GeV than the standard
calculations, after which they start to approach them again.
To quantify this difference I plot in Fig. \ref{plane}c the ratio:
\begin{equation}
R^{\mathrm{PMS}}(S) \equiv\frac{\sigma_{tot}^{\mathrm{NLO}}
(M_{sad},\mu_{sad})}
{\sigma_{tot}^{\mathrm{NLO}}(m_b,m_b,\overline{\mathrm{MS}})},
\label{PMSratio}
\end{equation}
together with the ratia
\begin{equation}
R^{\kappa}(S)
\equiv\frac{\sigma_{tot}^{\mathrm{NLO}}
(\kappa m_b, \kappa m_b,\overline{\mathrm{MS}})}
{\sigma_{tot}^{\mathrm{NLO}}(m_b,m_b,\overline{\mathrm{MS}})}
\label{CONVratio}
\end{equation}
for $\kappa=0.5$ and $\kappa=2$. The PMS-optimized scales thus lead to
predictions that are higher than those evaluated with the conventional
choice $\mu=M=m_b$. For low energies the former are close to those
corresponding to conventional calculations with $\kappa=0.5$, but at
high energies $R^{\mathrm{PMS}}(S)$ rises steeply, reaching its maximum
$R^{\mathrm{PMS}}(S)\doteq 2.2$ at about $\sqrt{S}=400$ GeV. At
$\sqrt{S}=1800$ GeV $R^{\mathrm{PMS}}(S)\simeq 1.9$, which is about the
same number as the factor characterizing the excess of D0 data over the
conventional NLO QCD predictions in $\overline{\mathrm{MS}}$, which
assume $\mu=M=m_b$. The fact that
the difference between the results based on PMS-optimized and
conventional choices of scales is a sensitive function of the collision
energy provides a way for checking the phenomenological relevance of the
optimization procedure based on the PMS idea of local stability.

The results presented above have been obtained using fixed order
perturbative QCD and could therefore be modified by the inclusion
of the resummation of small-$x$ and/or large $p_T$ logs. Note that
at the Tevatron the momentum fractions carried by partons
producing the $b\overline{b}$ pair are typically larger than
$0.006$, which is where the region of ``small-$x$'' usually
starts. So one might expect the approach based on unintegrated PDF
as used in CASCADE to improve the agreement with data, though it
is not obvious by how much as CASCADE uses only a LO hard
scattering cross sections. On the other hand, $R^{\mathrm{PMS}}$
peaks at $\sqrt{s}=400 $ GeV, for which the mean parton fractions
$\langle x\rangle \gtrsim 0.025$ are safely outside the small-$x$
region. Consequently, in the energy range where it is largest the
difference between standard NLO calculations in
$\overline{\mathrm{MS}}$ RS and PMS-optimized ones will likely
survive the effects of small-$x$ resummation.

The LO calculation of the energy dependence of mean $p_T$ of
the produced $b$-quarks, displayed by the dash-dotted curve in
Fig. \ref{plane}a, shows that it rises only very slowly
and approaches $7$ GeV at the Tevatron range. This indicates
that the resummation of large logs $\ln(p_T/m_b)$ may be important
for the tails of $p_T$ distributions, but cannot be expected to
influence significantly the results for the total cross section.

The results discussed so far were obtained with the GRV HO
parameterization of PDF of the proton. To check the dependence of the
above conclusions on the choice of PDF, all the calculations were
redone also with the MRS(R2) PDF used in \cite{d0b}. As illustrated in
Fig. \ref{plane}a, there is noticeable difference in the location of
saddle point at low energy, where the GRV saddle points are closer to
the ``diagonal'' ones than the MRS(R2) ones, but as the energy increase
both parameterizations give essentially the same results not only for
the for the positions of the saddle points, but
also for the energy dependence of the corresponding optimized total
cross sections. This is illustrated in Fig. \ref{plane}c, where
the ratio $R^{\mathrm{PMS}}$ is plotted for both
parameterizations. The conventional ratia $R^{1/2},R^{2}$ are so
close for both parameterizations, that only those corresponding to
GRV are displayed.

To get some feeling for the dependence of the above conclusions on the
mass of the $b$-quark, Figure \ref{plane} shows also the results
obtained for $m_b=11$ GeV, which is roughly the transverse mass
$m_T=\sqrt{m_b^2+p_T^2}$ at the Tevatron. Clearly, the difference
between standard and PMS-optimized calculations decreases with
increasing $m_b$ and the region of largest $R^{\mathrm{PMS}}$ shifts,
as expected, to larger energies.

\section{Summary and conclusions}
The results of this exploratory study demonstrate the sensitivity of the
NLO QCD calculations of the total cross section $\sigma_{tot}^{\mathrm{NLO}}
(\overline{b}b)$ to the choice of the renormalization and factorization
scales. In particular they show that the conventional assumption of
identifying these two scales leads at high energies
far away from the region of local stability as quantified by the Principle of
Minimal Sensitivity criterion. As a result, the conventional NLO calculations
yield results that are lower than the PMS-optimized ones, at the Tevatron
energies by a factor of about 2, which is approximately the factor needed to
bring the conventional calculations into agreement with the D0 and CDF data.

This observation cannot, however, be directly used as an explanation of
the excess of the Tevatron data over conventional NLO predictions because it
concerns the total cross section rather than the differential distribution
$\mathrm{d}\sigma/{\mathrm{d}}p_T$ measured by the D0 and CDF Collaborations.
The ongoing analysis \cite{srbek} of renormalization and factorization scale
dependence of such differential distribution indicates that the situation
for such distribution is similar to that discussed in this paper, with
additional dependence of the stability region on $p_T$ of the $b$-quark.
This paper will also address the question of the stability of the NLO QCD
calculations of the top quark production at both Tevatron and LHC.

I also don't claim that the choice of the renormalization and
factorization scale via the PMS-optimization explains fully the observed
discrepancy between data and conventional NLO QCD calculation of
$\sigma_{tot}^{\mathrm{NLO}}(\overline{b}b)$ or that it is the only effect
that might be responsible for such discrepancy. The purpose of this paper
is rather to point out yet another source of theoretical uncertainty of higher
order QCD calculations of quantities like
$\sigma_{tot}^{\mathrm{NLO}}(\overline{b}b)$ that should be properly
taken into account before drawing conclusions from the comparisons with
experimental data.

\acknowledgments
This works has been supported by the Ministry of Education of
the Czech Republic under the project LN00A006.

\end{document}